\documentclass[aps,prl,twocolumn]{revtex4}
\usepackage{amsmath,bm,epsfig}
\usepackage{color}

\def\Fbox#1{\vskip1ex\hbox to 8.5cm{\hfil\fboxsep0.3cm\fbox{%
  \parbox{8.0cm}{#1}}\hfil}\vskip1ex\noindent}  

\newcommand{\B}[1]{{\bm{#1}}}
\let \= \equiv \let\*\cdot \let\~\widetilde \let\-\overline

\def\d{{\rm d}}

\begin{document}
\title{Elastic Moduli in Nano-Size Samples of Amorphous Solids: System Size Dependence}
\author{Yossi Cohen and Itamar Procaccia}
\affiliation{Department of Chemical Physics, The Weizmann
 Institute of Science, Rehovot 76100, Israel}
\date{\today}
\begin{abstract}
This Letter is motivated by some recent experiments on pan-cake shaped nano-samples of metallic glass that indicate a decline in the measured shear modulus upon decreasing the sample radius. Similar measurements on  crystalline samples of the same dimensions showed a much more
modest change. In this Letter we offer a theory of this phenomenon; we argue that such results are generically expected for any amorphous solid, with the main effect being related to the increased contribution of surfaces with respect to bulk when the samples get smaller. We employ exact relations between the shear modulus and the eigenvalues of the system's Hessian matrix to explore the role of surface modes in affecting the elastic moduli.
\end{abstract}
\maketitle

{\bf Motivation}: This Letter is motivated by some experimental measurements of the elastic response of nanosamples of metallic glasses when the radius of cylindrically
shaped samples was reduced \cite{12Sam}.  In this
Letter we provide a theoretical explanation of this phenomenon. Since the experiments are performed at temperatures that are
much lower than the glass-transition temperature, we can in the following disregard thermal effect, and study the phenomenon
in athermal conditions.

\begin{figure}
\includegraphics[scale = 0.25]{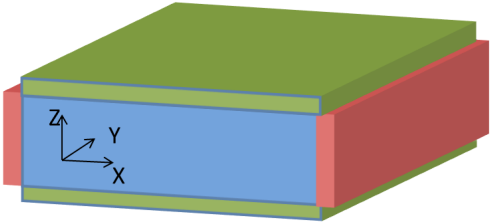}
\caption{(Color online) A cuboid shape. The top and bottom surfaces (green) were clamped to measure $\mu_{xz}$, and the opposite side wall (red) for $\mu_{xy}$.}
\label{shape}
\end{figure}
\begin{figure}
\includegraphics[scale = 0.60]{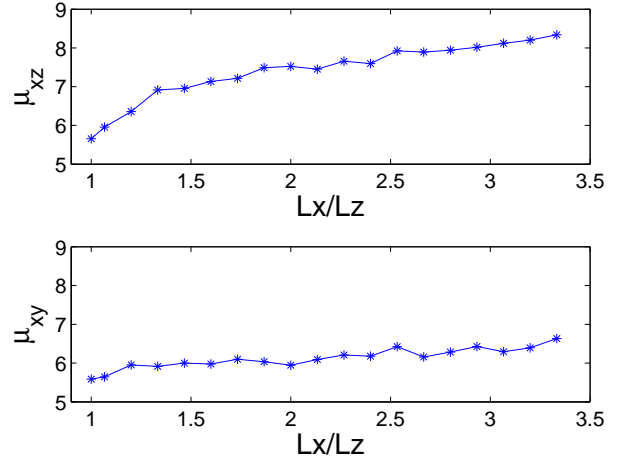}
\caption{(Color online) The dependence of the shear moduli $\mu_{xz}$ and $\mu_{xy}$ on the ratio of the edge size over the height for systems
varying by a factor of 3.5 in this ratio. One observes a change of about 50\% in $\mu_{xz}$ and about 15\% in $\mu_{xy}$.}
\label{ammod}
\end{figure}
\begin{figure}
\includegraphics[scale = 0.60]{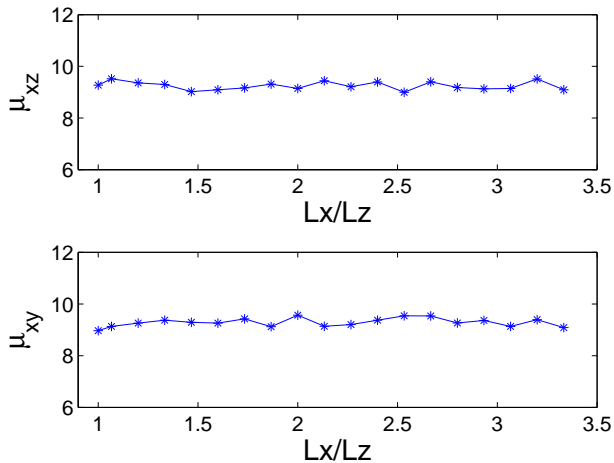}
\caption{(Color online) Same as Fig. \ref{ammod}, but with periodic boundary condition.}
\label{ammodPBC}
\end{figure}

{\bf Mathematical background}: For the sake of concreteness we study theoretically the shear modulus, stressing the difference in its exact calculation between a perfect crystalline sample and an amorphous solid sample of the same physical dimension. In both cases the
calculation of the shear modulus starts with the potential energy which for a shear-strained solid can be written as
$U(\{\B r_i(\gamma_{\alpha\beta})\},\gamma_{\alpha\beta})$ where $\{\B r_i\}_{i=1}^N$ are the positions of the $N$ particles and $\gamma_{\alpha\beta}$ is the applied strain.
(When possible we treat $\gamma_{\alpha\beta}$ as a scalar $\gamma$; all the equations can be written in tensorial form if required). In both cases we can
consider the deformation under an infinitesimal shear strain via the parameterized transformation on the particles coordinates
$\B J(\gamma) = I + \gamma \B h$ where $\B h$ determines the imposed deformation. Here comes the important difference between crystalline
and amorphous solids \cite{CL95}. For the former
\begin{equation}
\B r_i \to \B J\cdot \B r_i \ ,\quad \text{in a perfect crystalline solid}
\end{equation}
since the particles remain in mechanical equilibrium also after the deformation, with all the forces $\B f_i$ vanishing on each
and every particle. In an amorphous solid, on the other hand, even an infinitesimal deformation gives rise to non-zero forces
on the particles,  resulting in non-affine displacements $\B u_i$ that are necessary to restore mechanical equilibrium.
Thus for an amorphous solid
\begin{equation}
\B r_i \to \B J\cdot \B r_i +\B u_i\ , \quad\text{in an amorphous solid}.
\end{equation}
This crucial difference translates to a different calculation of the mechanical moduli; in a perfect crystal we can
write $d/d\gamma = \partial/\partial \gamma$. In an amorphous solid on the other hand
\begin{equation}
\frac{d}{d\gamma} =\frac{\partial}{\partial\gamma}+ \frac{d\B u_i}{d\gamma}\cdot \frac{\partial}{\partial \B u_i} =
\frac{\partial}{\partial\gamma}+\frac{d\B u_i}{d\gamma}\cdot\frac{\partial}{\partial\B r_i} \ . \label{rule}
\end{equation}
Applying this to the definition of the shear modulus, $\mu=(1/V) d^2 U/d\gamma^2$ (with $V$ being the volume) we find a different answer,
\begin{eqnarray}
\mu&=& \frac{1}{V}\frac{\partial^2 U}{\partial \gamma^2} \ , \quad \text{perfect crystalline solid}\ ,\\
\mu&=&\frac{1}{V}\left[\frac{\partial^2 U}{\partial \gamma^2}+\frac{\partial^2U}{\partial\gamma\partial\B r_i}\cdot \frac{d\B u_i}{d\gamma}\right]\ ,  \text{amorphous solid}\ . \label{amsol}
\end{eqnarray}
The shear modulus of the crystalline solid, which is the same as the first term in the shear modulus of the amorphous solid
is known as the Born approximation. This approximation is corrected by the second term which is due to the non-affine response
of the amorphous solid.

Eq. (\ref{amsol}) is brought to final form using the mechanical equilibrium condition $\d\B f_i/d\gamma=0$, which, using
Eq. (\ref{rule}) becomes $\partial \B f_i/\partial \gamma+(\partial \B f_i/\partial \B r_j)\cdot (d\B u_j/d\gamma)=0$ \cite{04ML}. Identifying
the force on the $i$th particle with $\B f_i \equiv -\partial U/\partial \B r_i$, we invert the last relation in favor of $d\B u_j/d\gamma$
to write
\begin{equation}
\frac{d\B u_i}{d\gamma} = -H^{-1}_{ij} \Xi_j \ ; \quad H_{ij} \equiv \frac{\partial^2 U}{\partial \B r_i \partial \B r_j} ; \quad \Xi_i\equiv
 \frac{\partial^2 U}{\partial \B r_i \partial \gamma} \ . \label{inverse}
\end{equation}
Using Eq. (\ref{inverse}) in Eq. (\ref{amsol}) we get the final result
\begin{eqnarray}
\mu&=& \frac{1}{V}\frac{\partial^2 U}{\partial \gamma^2} \ , \quad \text{perfect crystalline solid}\ ,\\
\mu&=&\frac{1}{V}\left[\frac{\partial^2 U}{\partial \gamma^2}-\B \Xi\cdot \B H^{-1}\cdot\B \Xi \right] \ , \text{amorphous solid} \label{amsolfin}
\end{eqnarray}

Noticing that the Hessian matrix $\B H$ is real and symmetric, the difference between the shear modulus of a crystalline and amorphous
solid is negative definite, necessarily reducing the shear modulus in the case of the amorphous solid compared to the crystalline
counterpart (with the same inter-particle potential). To understand the experimental observations of the mechanical softening
of smaller and smaller samples we need to understand why the correction term increases in absolute magnitude compared to the Born term
which is system-size independent to a very good approximation. We start by computing Eq. (\ref{amsolfin}) using numerical simulations.

{\bf Numerical Simulations}: Glassy amorphous samples were achieved by using a binary mixture of point particles interacting
via modified Lennard-Jones potential with three different characteristic interaction lengths $\sigma_{ss}=1$ and $\sigma_{\ell\ell}=1.4$  and
$\sigma_{s\ell}=1.18$. Details of the potentials can be found for example in Ref. \cite{11KLPZ,95KA}. Cuboid samples of fixed height ($\approx$ 15 particles) and square cross-section with varying edges (from 50 to 15 particles) were prepared by quenching from the melt using a gradient energy
method to cool the system to $T=0$ with zero pressure. The boundary conditions were free on all the edges except the two edges that were clamped to produce infinitesimal shear strain. Clamping the upper and lower edges we obtain a strain $\gamma_{xz}$ to measure $\mu_{xz}$ while clamping two opposite side walls resulted in measuring $\mu_{xy}$ (cf. fig. \ref{shape}). For any given system the Born term was computed directly from the partial derivatives \cite{10KLP,06ML}. The correction term was obtained by computing the Hessian matrix and the vector $\B \Xi$, again directly from their definitions, inverting the Hessian we compute $\B \Xi\cdot \B H^{-1}\cdot \B \Xi$ exactly. Results for both shear moduli as a function of the ratio
of width to height are shown in Fig. \ref{ammod}.
\begin{figure}
\includegraphics[scale = 0.55]{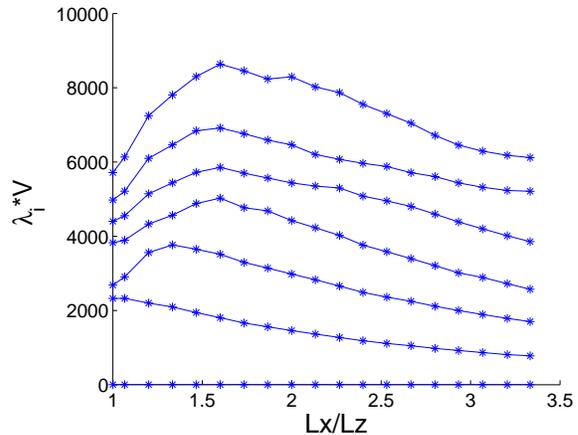}
\caption{(Color online) The seven lowest eigenvalues of the Hessian matrix multiplied by the volume of the sample, including the zero eigenvalue which relates to the Goldstone mode. Note the strong decline as $Lx/Lz$ approaches $1$. }
\label{eigam}
\end{figure}
\begin{figure}
\includegraphics[scale = 0.55]{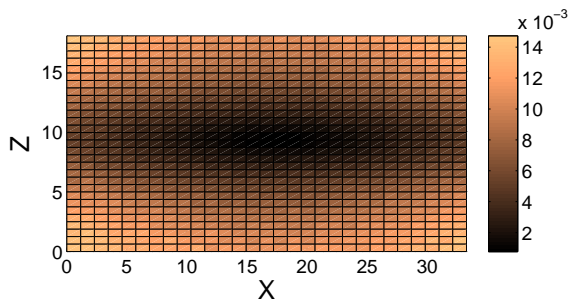}
\includegraphics[scale = 0.58]{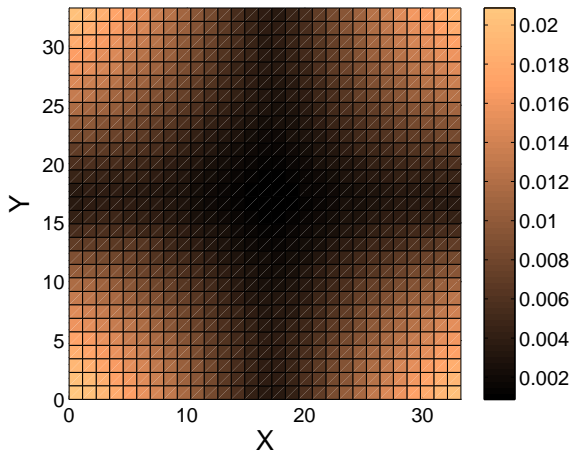}
\caption{(Color online) Upper panel: an $xz$ projection by averaging over the $y$ position of the magnitude of the elements of the first
nontrivial eigenfunction of the Hessian for the case $L_x/L_z = 1.86$. Lower panel: the same but for an $xy$ projection, averaging
over $z$. The color code is given on the right. Note the strong concentration on the surfaces }
\label{eigenfun1}
\end{figure}
\begin{figure}
\includegraphics[scale = 0.55]{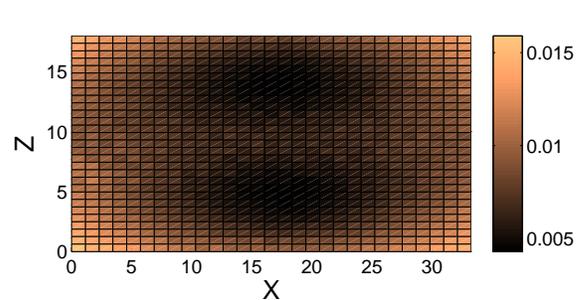}
\includegraphics[scale = 0.58]{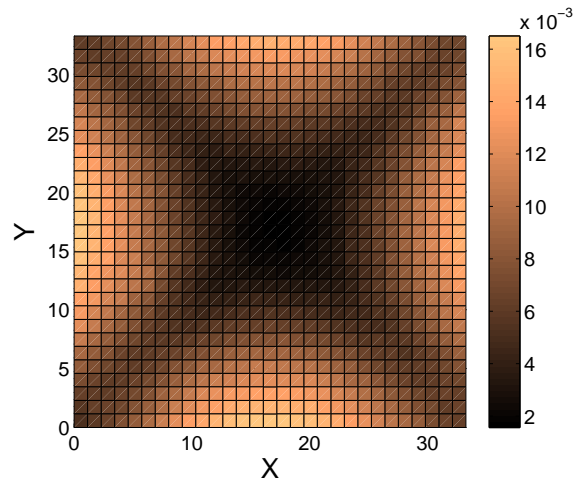}
\caption{(Color online) The same as in the previous figure, but for the second nontrivial eigenfunction of the Hessian. Again we see that the
lowest eigenvalues belong to eigenfunctions that are strongly concentrated on the surfaces.}
\label{eigenfun2}
\end{figure}

We observe a much larger change in $\mu_{xz}$ (about 50\%) than in $\mu_{xy}$ (about 15\%). We will argue below that this difference
stems from the contribution of free surfaces. In order to emphasize the role of the free surfaces, we performed the same measurements of the shear modulus in an infinite system made of finite cells which repeat by applying periodic boundary conditions. In this case the shear modulus exhibits the value of the bulk modulus for both directions and for various cell size, cf. Fig. \ref{ammodPBC}.

{\bf Theoretical explanation}: Having observed a tendency towards softening that is very comparable in magnitude to the
experimental one, we are now in a position to provide a theoretical explanation. Qualitatively speaking we expect that particles
close to the free surfaces should have softer fluctuations in their positions, being less constrained than particles in the
bulk. By lowering the volume of the samples we make the contribution of particles at and near the surface more dominant, resulting
in an overall softening of the mechanical response. Being equipped with the theory presented above, we can prove this
qualitative intuition by examining the eigenvalues and eigenfunctions of the Hessian matrix $\B H$ and how they depend on the volume.
The negative definite term in the amorphous shear modulus can written in a way that makes the contribution of the eigenvalues
of the Hessian explicit by expanding $\B \Xi $ in the eigenfunctions $\B \Psi^{(k)}$ of $\B H$:
\begin{equation}
\B \Xi = \sum_k a_k \B \Psi^{(k)}\ ; \quad a_k\equiv \B \Xi\cdot \B \Psi^{(k)} \ .
\end{equation}
With this in mind we can write the correction term to the Born approximation in the form
\begin{equation}
-\frac{1}{V}\B \Xi\cdot \B H^{-1}\cdot \B \Xi = \frac{1}{V}\sum_k \frac{|a_k|^2}{\lambda_k} \ ,
\end{equation}
where $\lambda_k$ is the eigenvalue of the Hessian matrix associated with the eigenfunction $\B \Psi^{(k)}$.

In Ref. \cite{11HKLP} it was shown that $|a_k|^2$ is roughly system size independent. This is not the case for the eigenvalues.
In Fig. \ref{eigam} we show the seven lowest eigenvalues of the Hessian matrix (including the Goldstone mode \cite{CL95}) multiplied by
the volume, and observe the strong decline in these products when $L_x/L_z$ becomes smaller than, say, 1.6. This decline will
increase the negative correction term and accordingly will reduce the moduli. It is also obvious why $\mu_{xz}$ is much more
sensitive to this effect than $\mu_{xy}$ - in the former case, the large top and bottom surfaces are clamped and only the side walls are free to contribute. Upon increasing the ratio $Lx/Lz$ the contribution of the free surfaces decreases, and the shear modulus increases until we reach the value of the bulk. In the latter case, the large top and bottom surfaces are free,  contributing to the softening of the system. Clearly, we need large free
surfaces to observe a sizeable effect; when we reduce the volume the ratio of surface to volume increases, lowering the overall shear
modulus. We can directly prove
that the effect is strongly connected to the free surfaces by examining the eigenfunctions associate with the lowest lying eigenvalues.
In Figs. \ref{eigenfun1} and \ref{eigenfun2} we present the first and second eigenfunction by showing the magnitude of elements of the
eigenfunctions in real space.

\begin{figure}
\includegraphics[scale = 0.60]{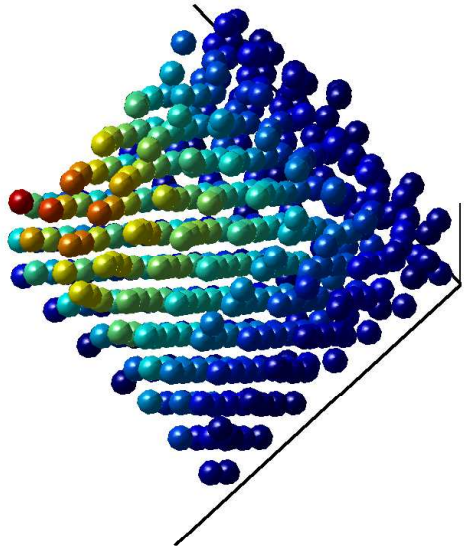}
\caption{(Color online) An example of a crystalline solid in three dimensions. The order of the crystal is distorted due to free surface effect. Note that the blue particles are on the surface and as the color code tends to red we observe bulk crystalline order.}
\label{solid}
\end{figure}
\begin{figure}
\includegraphics[scale = 0.60]{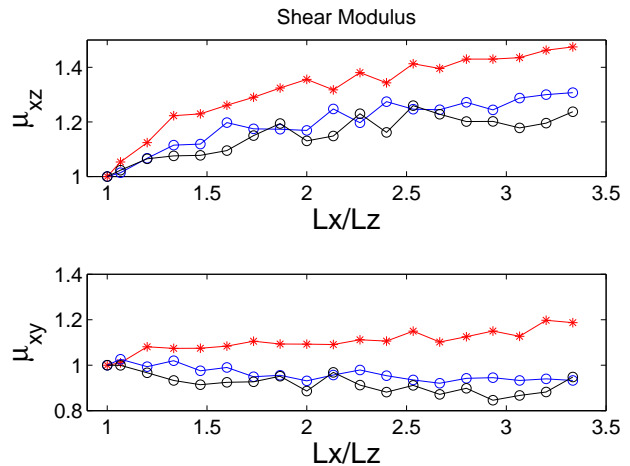}
\caption{(Color online) The dependence of the shear moduli on the ratio of the edge size over the height, for an amorphous solid (stars), and a crystalline solid (circles) contains either small or large particles, normalized to the value of the shear modulus of a cube.}
\label{comparison}
\end{figure}

\begin{figure}
\includegraphics[scale = 0.50]{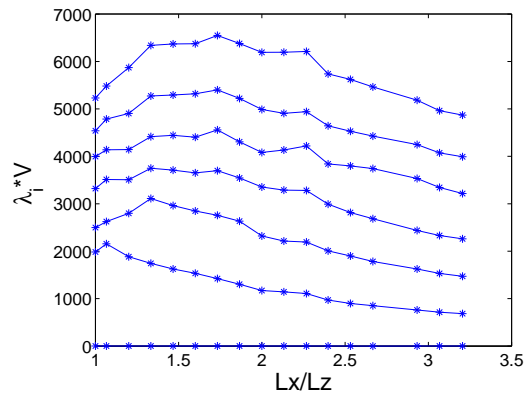}
\caption{(Color online)The same as in Fig. \ref{eigam} but for a crystalline solid. Here the decline is weaker than the case of amorphous solid.}
\label{eigsol}
\end{figure}
We see that both eigenfunctions are strongly concentrated on the surfaces. This is a direct evidence that the lowest eigenvalues
of the Hessian, which are responsible for the softest mechanical response, indeed belong to modes that live on the surfaces of
our samples as proposed above.

At this point it is interesting to compare the simulation results to similar simulations on crystalline samples
of the same range of sizes and the same geometry. We attempted to produce perfect crystalline samples by
arranging either `small' or `large' particles on a lattice at zero temperature. We find that when we minimize
the energy, surface tension effects destroy the crystalline order near the interfaces, leaving us with bulk crystals
with amorphous interfaces, as seen for example in Fig. \ref{solid}.

We thus expect to still see an effect of
softening when the surface to bulk ratio is reduced, but the effect should be smaller compared to the completely
amorphous samples. For the sake of comparison we show normalized values of $\mu_{xz}$ (normalized to the
lowest value) for the three types of samples, amorphous and crystalline with small or large particles. In Fig. \ref{comparison} we see that the crystalline samples show very similar curves for $\mu_{xz}$, and the effect
is considerably smaller than for the amorphous solid. The relative decrease in the size of the effect is also
born out by the values of the lowest eigenvalues of the Hessian matrix (multiplied by the volume) as seen in
Fig. \ref{eigsol}. The eigenvalues times the volume still have a reduction for the largest values of the surface
to volume ratio, but the degree of reduction is considerably smaller, cf. Fig. \ref{eigam}. We expect that for larger samples (as in the experiment) where the surface to bulk ratio is smaller than in our simulations, the effect will be even smaller than in our present simulations.

In summary, we have shown that the main reason that is responsible for the softening of the mechanical response of
nano-samples as a function of their size comes from the softer normal modes of the hessian matrix that are concentrated
near the boundaries. These reflect the physics of increased freedom of particles near the surfaces compared to the
highly constrained particles in the bulk.

\acknowledgments
This work had been supported in part by an advanced ``ideas" grant of the European Research Council, the Israel Science
Foundation and the German Israeli Foundation.

\end{document}